\documentclass[12pt]{iopart}

\usepackage{bm}
\usepackage{graphicx}
\usepackage{color}

\begin{document}

\title[The sign of electron $g$-factor in
GaAs$_{1-x}$N$_{x}$ measured by using the Hanle effect]{The sign
of electron $g$-factor in GaAs$_{1-x}$N$_{x}$ measured by using
the Hanle effect}

\author{V K Kalevich, E L Ivchenko, A Yu Shiryaev,
M M Afanasiev and A Yu Egorov}
\address{A.F.~Ioffe Physico-Technical Institute, St.
Petersburg 194021, Russia}

\author{M Ikezawa and Y Masumoto}
\address{Institute of Physics, University of Tsukuba, Tsukuba
305-8571, Japan}

\ead{kalevich@solid.ioffe.ru}

\begin{abstract}
Positive signs of the effective $g$-factors for free electrons in the conduction band
and electrons localized on deep paramagnetic centers have been measured in nitrogen
dilute alloy GaAs$_{0.979}$N$_{0.021}$ at room temperature. The $g$-factor signs have
been determined from an asymmetry in the depolarization of edge photoluminescence in a
transverse magnetic field (Hanle effect) at the oblique incidence of the exciting
radiation and oblique-angle detection of the luminescence. The tilted spin polarization
of free electrons is induced under interband absorption of circularly polarized light,
and the paramagnetic centers acquire spin polarization because of spin-dependent capture
of free spin-polarized electrons by these centers. The measured Hanle curve is a
superposition of two lines, narrow and broad, with the widths $\sim400$\,G and
$\sim50000$\,G, arising due to the depolarization of localized and free electrons,
respectively. The difference in the linewidths by two orders of magnitude strongly
indicates much longer spin lifetime for the paramagnetic centers as compared with that
for the free carriers. The magnitude and direction of the asymmetry in the measured
Hanle curve have been found to depend on the partial contributions to the recombination
radiation from the heavy- and light-hole subbands split by a uniaxial deformation of the
GaAs$_{1-x}$N$_{x}$ film grown on a GaAs substrate. We have extended the theory of
optical orientation in order to calculate the excitation spectrum of the photoelectron
tilted-spin polarization and the circularly-polarized luminescence spectrum taking into
account that, in the strained samples under study, the light-hole subband lies above the
heavy-hole one. The results have further been used to calculate the shape of Hanle curve
as a function of the excitation and registration energies as well as the incidence and
detection angles and to compare the theory with experiment.
\end{abstract}

\maketitle

\section{Introduction}
Dilute III-N-V alloys with the group-V cations partially substituted by nitrogen, e.g.,
GaAsN and InGaAsN, have recently attracted particular attention owing to a number of
their unusual properties. Thus, an increase of nitrogen content in the alloy is
accompanied by an anomalous reduction of the fundamental band gap, by more than 0.1\,eV
per  1\% of N as the nitrogen concentration increases from zero up to 4\%~
\cite{AndoJJAP1992,TuAPL1977,KondowJCG1998,EgorovJAP2005}, as well as by a drastic
increase of the electron effective mass and enhanced nonparabolicity of the $\Gamma_6$
conduction band \cite{SkierbAPL2000,Buaynova2000,SkierbPRB2001, italo-irlandPRB2006}
(for reviews, see also \cite{N-book2005,N-book2008}). These specific properties are
related to the N-substitution-induced resonant electronic states within the conduction
band continuum and an anticrossing of the localized nitrogen states with the extended
conduction states of the semiconductor matrix
\cite{WalukPRL1999,ZungerPRB2001,ReilyPRL2004}. Qualitatively, the available
experimental data are consistent with the simple phenomenological band anticrossing
(BAC) model which takes into account the anticrossing hybridization only with the levels
of isolated (single) nitrogen atoms \cite{WalukPRL1999}. As a result, the two hybridized
conduction bands, $E_{+}$ and $E_{-}$, are formed. As the nitrogen concentration
increases, the lower band, $E_{-}$, shifts down in energy reducing the fundamental
energy-band gap $E_{g}$. Simultaneously, the band-bottom effective mass becomes heavier
and the nonparabolicity coefficient rises considerably.

The band-structure modification involves a remarkable change in the gyromagnetic factor,
$g$, of electrons in the conduction band $E_{-}$~\cite{PolyakiPRB2005,PettinariPRB2006}.
The change is caused by two reasons. Firstly, as in the case of conventional alloys
In$_{x}$Ga$_{1-x}$As \cite{Herm&WeisbuchPRB1977, H&W_OO}, the band-gap shrinkage leads
to a monotonous downward shift of the effective $g$-factor below the value of
\emph{g}-factor in GaAs matrix ($g=-0.44$ in GaAs at helium temperature)
\cite{PolyakiPRB2005}. Secondly, the coupling with isolated nitrogen levels, with the
Land\'{e} factor being positive and close to the electron $g$-factor in vacuum
$g_{0}=2$, tends to shift the \emph{g} value upwards, as it has been first found
experimentally in InGaAsN when studying the dependence of the absorption coefficient on
the magnetic field \cite{PolyakiPRB2005}. The two effects have opposite signs and
partially compensate each other. The BAC model predicts an initial growth in the
dependence $g(x)$ \cite{PolyakiPRB2005}. The dependence is smooth and, as calculated for
GaAs$_{1-x}$N$_{x}$ in \cite{PettinariPRB2006}, it has a maximum value $g\approx0$ at
$x\sim1\%$. At the same time, detailed measurement and comparison of the left- and
right-circularly polarized photoluminescence spectra performed in GaAs$_{1-x}$N$_{x}$ in
a longitudinal magnetic field at helium temperature for $x\leq0.6\%$ have shown that
\emph{g} grows sharply with increasing $x$, reverses its sign from negative to positive
at $x\approx0.04\%$, reaches a value of 0.7 in the range $x=0.04 \div 0.1\%$ and retains
that value with small deviations as $x$ increases up to $0.6\%$ \cite{PettinariPRB2006}.
Such an abrupt dependence of $g$ on $x$ at small $x$ has been interpreted in terms of a
modified $\textbf{k}\cdot\textbf{p}$ model \cite{ReilyPRL2004, PettinariPRB2006}. This
model takes into account the hybridization of the host conduction band not only with the
localized levels due to single N atoms but also with the levels which are formed by
clusters of N atoms, pairs and triplets, and lie closer to the conduction-band bottom.

The data on the $g$-factor in GaAs$_{1-x}$N$_{x}$ with high
nitrogen concentration are highly desirable for the purpose of
checking various theoretical models. However, to the best of our
knowledge, such experimental data for $x>0.6\%$ are lacking.

In the present work the sign of the electron $g$-factor in the
GaAs$_{1-x}$N$_{x}$ conduction band has been measured at room
temperature for $x=2.1\%$. To this end, we used the dependence of
the sign of the asymmetry of Hanle effect of optically
spin-polarized electrons, measured under oblique excitation, on
the $g$-factor sign. This method is based on the relationship
between the direction of Larmor precession of electron spin in a
magnetic field and the $g$-factor sign. As compared to other
methods, this method is advantageous for studying crystals with
high nitrogen content, particularly at high (up to room)
temperatures when the photoluminescence occurs over a broad
spectrum. The fundamentals of the method are presented in Section
3, together with the discussion of peculiarities of the electron
spin polarization and the oblique Hanle effect caused by the
splitting of the $\Gamma_8$ valence band due to a uniaxial strain
of the GaAsN films grown on GaAs substrates.

Section 4 provides the theory extended in order to calculate the magnitude and
orientation of photoelectron spin polarization, the circular polarization of
photoluminescence and the Hanle-curve shape as a function of the energy of absorbed and
emitted photons in a uniaxially strained semiconductor with the top of the light-hole
subband to be higher in energy than the top of the heavy-hole subband.

In the GaAsN alloys under study, the optical spin polarization of conduction electrons
at room temperature has been found to be anomalously
high~\cite{EgorovJAP2005,JETPLett2005}. This is the result of spin-dependent
recombination of polarized conduction electrons on deep paramagnetic centers which
apparently appear during the introduction of nitrogen atoms into GaAs. The capture of
free electrons is accompanied by efficient dynamic spin polarization of electrons bound
on the centers. In its turn, the polarized centers kinetically affect the polarization
of free electrons so that a strongly coupled spin system of free and localized electrons
is formed. A change in the polarization of localized electrons manifests itself in a
remarkable change of the free-electron polarization. Thus, from the analysis of the
curve of magnetic depolarization of the interband photoluminescence contributed by free
particles, one can determine the $g$-factor signs of both free and bound electrons. In
Section 5 we present experimental results together with the results of numerical
calculation taking into account the influence of spin-dependent recombination on the
electron spin polarization.

\section{Samples and experimental details}
We studied the undoped $0.1\,\mu$m-thick GaAs$_{0.979}$N$_{0.021}$
layer grown by rf-plasma-assisted solid-source molecular-beam
epitaxy at 350--450$^{\circ}$C on semi-insulating (001) GaAs
substrate \cite{EgorovJAP2005}. The nitrogen content and
crystallinity of the grown layer were examined by x-ray
diffraction technique. The as-grown structure was annealed for
5\,min at 700$^{\circ}$C in a flow of arsenic in the growth
chamber. Continuous-wave tunable Ti:sapphire laser was used for
photoluminescence (PL) excitation. Spin polarization of electrons
was created upon the interband absorption of circularly polarized
light \cite{OO}. It was monitored by measuring the degree of
circular polarization of PL, defined as $\rho = (I^+ - I^-)/(I^+ +
I^-)$, where $I^+$ and $I^-$ are the right ($\sigma^{+}$) and left
($\sigma^{-}$) circularly polarized PL components. The value
$\rho$ and PL intensity in a wavelength range up to 1.4\,$\mu$m
were measured using a high-sensitive polarization analyzer
\cite{Kulkov} comprising a quartz polarization modulator
\cite{Jasperson}, a lock-in two-channel photon counter, and a
photomultiplier with an InGaAsP photocathode. The measurements
were carried out at 300\,K.

\begin{figure}
\includegraphics[width=0.5\linewidth]{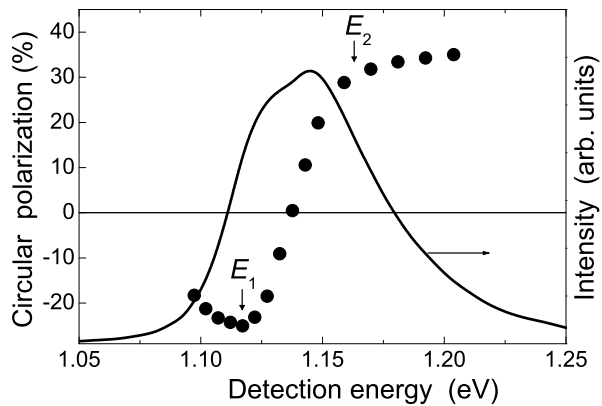}
\caption{\label{PL^Ro(det)} Spectral dependencies of the PL
intensity (solid curve) and of the PL circular polarization degree
(circles) for the GaAs$_{0.979}$N$_{0.021}$ layer under normal
incidence ($\theta = 0$) of the pump beam and detection of the PL.
Excitation energy $\hbar\omega_{\rm exc}=1.312$\,eV. Arrows
indicate the PL detection energies $E_{1}=1.117$ and $E_{2}=1.163$
at which the Hanle curves presented in Fig.~2 are measured. }
\end{figure}

Figure\,\ref{PL^Ro(det)} shows the spectra of PL intensity (solid curve) and of PL
circular polarization degree (circles) measured in GaAs$_{0.979}$N$_{0.021}$ layer under
normal incidence of the pump beam and the detection of luminescence along the growth
axis. The PL spectrum consists of two strongly overlapping inhomogeneously broadened
bands where the low-energy PL band is negatively polarized (relative to the polarization
of the exciting beam), whereas the high-energy band is polarized positively. As we have
shown earlier \cite{EgorovJAP2005}, the two PL bands appear due to splitting of the
light- and heavy-hole valence subbands induced by uniaxial compression along the growth
axes of GaAs$_{1-x}$N$_{x}$ film grown on GaAs substrate. In its turn, this compression
arises from large lattice mismatch between the film and substrate. The crystal splitting
of the PL bands, $\Delta_{c}$, increases with increase of the nitrogen content and at
$x=2.1\%$ equals 29\,meV. An interrelation of the sign of the PL circular polarization
and the type of recombination transition in GaAs$_{1-x}$N$_{x}$ is considered in detail
in \cite{EgorovJAP2005}. Under simultaneous excitation of electrons from both valence
subbands, realized in Fig.\,\ref{PL^Ro(det)}, the negative and positive PL polarizations
are due to recombination of conduction electrons with the light and heavy holes,
respectively. Since under an uniaxial compression the top of the light-hole subband is
situated above the top of the heavy-hole subband \cite{BirPikus}, the low-energy PL band
is polarized negatively.

In bulk semiconductor, the spin relaxation rate of conduction electrons grows
dramatically with temperature and decreases electron polarization down to units and
fractions of $\%$ at room temperature \cite{OO}. At the same time, the absolute values
both of negative and positive polarization in Fig.\,\ref{PL^Ro(det)} reach $30 \div
35\%$, which is near to the maximal magnitude of $50\%$ determined by selection rules.
Such anomalous enhancement of free electron polarization in GaAsN is due to dynamic
polarization of electrons bound on deep paramagnetic centers
\cite{JETPLett2005,Toulouse2007,JETPLett2007}. The paramagnetic centers arise with an
incorporation of nitrogen in GaAs and are polarized as a result of spin-dependent
capture on them of the polarized conduction electrons. The polarized centers increase
the free electron polarization which can reach $\approx100\%$ at strong pumping.

\section{The Hanle effect in GaAsN alloys}
\subsection{Electron Hanle effect under normal excitation and detection}
\begin{figure}
\includegraphics[width=0.45\linewidth]{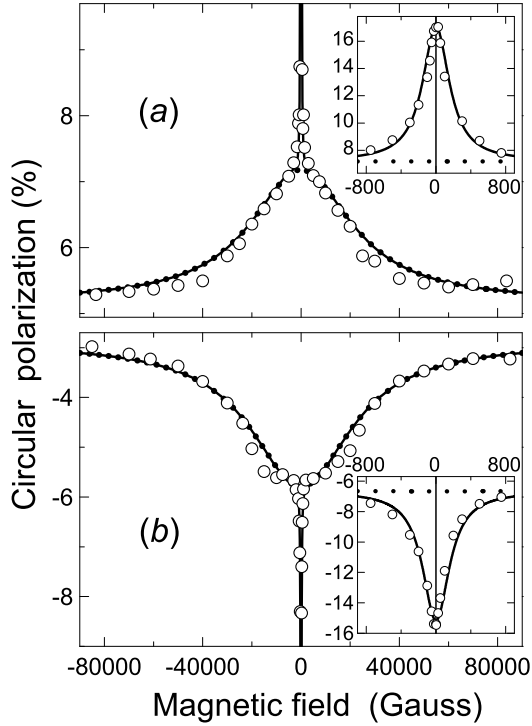}
\caption{\label{0teta} Experimental Hanle curves (open circles)
measured in GaAs$_{0.979}$N$_{0.021}$ at excitation and detection
along the normal to the sample. Excitation energy
$\hbar\omega_{\rm exc}=1.312$\,eV, detection energy
$\hbar\omega_{\rm det}$ is equal to
$\emph{E}_{2}=1.163$\,eV~(\emph{a}) and
$\emph{E}_{1}=1.117$\,eV~(\emph{b}). Solid curves are
superpositions of the calculated Hanle curves for free (dotted
curves) and bound electrons (see text for details). Inserts show
the initial parts of the Hanle curves in the range of small
magnetic fields. }
\end{figure}

The Hanle effect is a depolarization of photoluminescence by a magnetic field  ${\bm B}$
directed perpendicular to continuous-wave pump beam \cite{OO}. The effect originates
from Larmor precession of electron spins, which destroys their polarization. In the
simplest case, Hanle effect is described by Lorentzian
$\rho(B)/\rho(B=0)=1/(1+B^{2}/B_{1/2}^{2})$ with half-width at half maximum,
$B_{1/2}=\hbar/g\mu_B T_s$, where $g$ is the electron Land\'{e} $g$-factor, $\mu_B$ is
the Bohr magneton, and $T_s$ is the electron spin-polarization lifetime.
Figure\,\ref{0teta} shows the Hanle curves measured in GaAs$_{0.979}$N$_{0.021}$ for the
positive (a) and negative (b) PL polarization under the normal incidence of pump light
onto the sample and the detection of the PL in the opposite direction (backscattering
configuration). The measurements were carried out with the energy of the excitation
quantum $\hbar\omega_{\rm exc}=1.312$\,eV, at which the conduction electrons are excited
from both valence subbands. One can see that for both positive and negative PL
polarization the Hanle curve is superimposition of the narrow and wide curves with the
half-widths being different by two orders of magnitude. A qualitative explanation of
such a complex shape of the $\rho(B)$ dependency is the following. Spin lifetime of
electrons bound on paramagnetic centers, $T_{sc}$, can exceed spin lifetime of free
electrons, $T_s$, by orders of magnitude at room temperature. Therefore, the half-width
of Hanle curve of bound electrons, $B_{1/2}^c=\hbar/g_{c}\mu_B T_{sc}$ (here $g_c$ is
the gyromagnetic \emph{g}-factor of bound electrons), should be far less than the
half-width of Hanle curve of free electrons $B_{1/2}$. As noted above, the
spin-dependent recombination results in formation of a strongly coupled spin-system of
free and bound electrons, where a variation of the polarization of the centers is
accompanied by change in the free-electron polarization. Due to this, the $\rho(B)$
curve is the sum of two curves with strongly different half-widths, where the narrow
curve describes the Hanle effect of bound electrons and the wide curve presents
depolarization of free electrons. Solid curves in Fig.\,\ref{0teta} are calculated using
the equation $\rho(B)=\rho_{0}/[1+(B/B_{1/2})^{2}]+
\rho_{0c}/[1+(B/B_{1/2}^{c})^2]+\rho_{\rm res}$ at $B_{1/2}=25000$\,G and
$B_{1/2}^{c}=185$\,G. It is seen that the calculated curves describe reasonably the
measured ones in Fig.\,\ref{0teta} for both positive and negative polarization. During
the calculation the fitting parameters were $\rho_{0}$, $\rho_{0c}$ and $\rho_{\rm
res}$. At present, the origin of the constant polarization $|\rho_{\rm res}| \sim4\%$ is
not clear and additional studies should be done for its elucidation.

\subsection{Electron Hanle effect under oblique excitation and detection}
\begin{figure}
\includegraphics[width=0.25\linewidth]{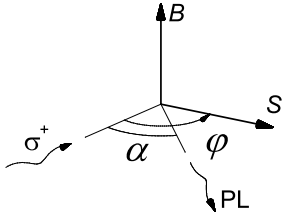}
\caption{\label{MirrorConfig} Experimental configuration for the
$g$-factor sign measurements. The angle $\alpha$ is the angle
between the excitation beam and the direction of the detected
photoluminescence.}
\end{figure}

The Hanle effect can provide measuring the sign of the Land\'{e} $g$-factor of electrons
\cite{Vekua1974,OptSpect1982,FTT1997} since a rotation direction of mean electron spin
in a magnetic field depends on the $g$-factor sign. The most suitable way to find the
$g$-factor sign is to use a specular geometry of experiment. In this geometry, the
exciting light falls on the sample obliquely, the luminescence is detected at an angle
to the exciting beam while a magnetic field is perpendicular to both the excitation and
detection directions, and lies in the crystal surface plane \cite{OptSpect1982}, as
shown in Fig.\,\ref{MirrorConfig}.

In specular geometry, the dependence $\rho(B)$ in bulk zinc-blende semiconductor has the
form \cite{OptSpect1982}:
\begin{equation}
\label{HanleBulk} \frac{\rho(B)}{\rho_0}=\frac{\cos \alpha +
\varphi \sin \alpha}{1+\varphi^2}~~,
\end{equation}
where $\alpha$ is the angle between the directions of excitation and detection, $\rho_0$
is the degree of polarization for $B=0$ and $\alpha=0$, $\varphi=\Omega T_s=g\mu_B
BT_s/\hbar$ is the angle through which the mean spin $\bm{S}$ of electrons with the
\emph{g}-factor equal to \emph{g} turns during the electron spin lifetime
$T_s=\tau\tau_s/(\tau+\tau_s)$, $\tau$ and $\tau_s$ are the electron lifetime and spin
relaxation time in the conduction band, $\bm{\Omega}=g\mu_B\bm{B}/\hbar$ is the Larmor
frequency, and the Born magneton $\mu_B>0$. Equation\,(\ref{HanleBulk}) was derived
taking into account the fact that in a bulk unstrained GaAs-type crystal the mean
electron spin in the moment of creation, $\bm{S}_0$, is parallel to the exciting beam,
and the degree of PL polarization is equal to the projection of $\bm{S}$ on the
direction of detection: $\rho = \bm{S}\bm{n}_1$, where $\bm{n}_1$ is the unit vector
along the direction of registration \cite{OO,DP1973}. At $\alpha=0$, the Hanle curve is
a symmetrical function on magnetic field as, for example, one can see in
Fig.\,\ref{0teta}. For $\alpha\neq0$, the $\rho(B)$ plot is not symmetrical with respect
to the $B=0$ point. The value of $\rho$ is larger for the magnetic field direction at
which the electron spin precesses toward the luminescence observation axis. Note, that
for measuring the \emph{g}-sign it is sufficient to record the asymmetry in the
dependence of the absolute value of $\rho$ on $B$.

The above method acquires distinctive features \cite{FTT1997} if
the heavy and light hole subbands are split due to the uniaxial
deformation in a bulk crystal or due to quantum-size effect in a
nanostructure.

\begin{figure}
\includegraphics[width=0.3\linewidth]{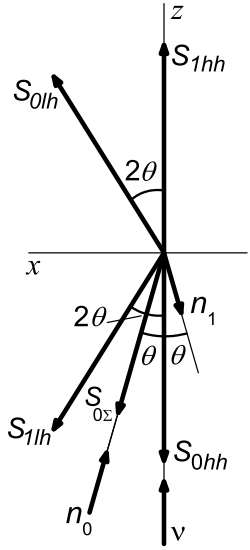}
\caption{\label{Ezhik} Scheme of orientation of spins ${\bm
S}_{0hh}$, ${\bm S}_{0lh}$, ${\bm S}_{0\Sigma}$ and vectors ${\bm
S}_{1hh}$, ${\bm S}_{1lh}$ with respect to the directions ${\bm
n}_0$ of excitation, ${\bm n}_1$ of detection and ${\bm \nu}$ of
deformation (${\bm \nu}$ is a perpendicular to the sample surface)
in the case of specular configuration.}
\end{figure}

First, in this case the direction of the mean spin of
photogenerated electrons ${\bm S}_0$ may not coincide with the
direction of the exciting beam. Theory \cite{DP1973} (see also
\cite{DPoo}) developed for uniaxially strained crystals yields the
following expressions for the mean spins of the electrons created
with ${\bm k}\approx0$ from states close to the tops of light-hole
($lh\to c$ transition) and the heavy-hole ($hh\to c$ transition)
subbands:
\begin{equation} \label{shhslh}
\label{k0} {\bm S}_{0lh}=\frac{3({\bm \nu} {\bm n}_0) {\bm \nu} -
2 {\bm n}_0}{5-3({\bm \nu} {\bm n}_0)^2},~~~~~~~~~~~
 {\bm S}_{0hh}=\frac{- {\bm \nu}({\bm \nu}
{\bm n}_0)}{1+({\bm \nu} {\bm n}_0)^2}~~,
\end{equation}
where ${\bm \nu}$ and ${\bm n}_0$ are unit vectors along the deformation (growth) axis
and pump beam direction. As seen from Eqs.\,(\ref{k0}), the spin ${\bm S}_{0hh}$ is
parallel to the deformation axis for any light incidence angle, whereas the angle
between ${\bm S}_{0lh}$ and ${\bm \nu}$ is twice the angle $\theta$ between the
excitation direction and the growth axis (Fig.\,\ref{Ezhik}) \cite{note1}. The
light-hole and heavy-hole splitting can be neglected if the energy of the photogenerated
holes exceeds substantially the splitting energy $\Delta_c$ \cite{DP1973}. Such
situation can be realized when $\hbar\omega_{\rm exc}-E_{g}\gg\Delta_c$, where $E_{g}$
is the forbidden gap of the deformed crystal. In this case, the electrons are excited
from both valence subbands, and, as in a bulk unstrained crystal, their mean spin ${\bm
S}_{0\Sigma}$ is directed along the exciting beam: ${\bm S}_{0\Sigma}\downarrow\uparrow
{\bm n}_0$ (Fig.\,\ref{Ezhik}). Thus, with the increase of excitation energy from
$\hbar\omega_{\rm exc}\approx E_{g}$ towards $\hbar\omega_{\rm exc}\gg E_{g}+\Delta_c$,
the spin ${\bm S}_{0}$ changes from ${\bm S}_{0lh}$ to ${\bm S}_{0\Sigma}$ rotating
counterclockwise through angle ($180^{\circ}-3\theta$).

Second, in the case of valence-subband splitting, the PL
polarization is determined by projection of ${\bm S}$ not on the
direction of observation ${\bm n}_1$, as this should be for a bulk
unstrained semiconductor, but rather on the vector ${\bm S}_1$,
which depends on the angle between ${\bm n}_1$ and ${\bm \nu}$
vectors and on the actual recombination energy \cite{DP1973,OO}:
\begin{equation} \label{e7}
\rho=-4{{\bm S}{\bm S}_1}~.
\end{equation}
For recombination with the tops (${\bm k} \approx 0$) of the
light-hole and heavy-hole subbands, the vectors ${\bm S}_{1lh}$
and ${\bm S}_{1hh}$ are given by the same expressions (\ref{k0})
as ${\bm S}_{0lh}$ and ${\bm S}_{0hh}$, but with ${\bm n}_0$ being
replaced by ${\bm n}_1$. We readily see that ${\bm S}_{1hh}$ is
always parallel to the growth axis, while ${\bm S}_{1lh}$ is at an
angle $2\theta$ to this axis (Fig.\,\ref{Ezhik}). If the PL
detection energy $\hbar\omega_{\rm det}\approx E_{g}$,
recombination with the light-holes is only possible and,
therefore, ${\bm S}_{1}= {\bm S}_{1lh}$. When $\hbar\omega_{\rm
det}-E_{g}\gg\Delta_c$, the crystal valence-band splitting can be
ignored and an effective direction of the PL detection, given by
vector ${\bm S}_1$, coincides with the real one: ${\bm S}_1
\downarrow\uparrow {\bm n}_1$. So, with increase of the detection
energy, the vector ${\bm S}_1$ changes its direction from ${\bm
S}_{1lh}$ to $(-{\bm n}_1)$, rotating clockwise through angle
($180^{\circ}-3\theta$).

Since both the value and direction of the ${\bm S}_0$ and ${\bm S}_1$ vectors are energy
dependent, the Hanle curves will be different for different excitation and detection
energies. However, as shown in \cite{FTT1997}, they can still be described by
Eq.\,(\ref{HanleBulk}), provided the angle $\alpha$ is replaced by the angle
$(\gamma-\beta)$, where $\gamma$ and $\beta$ are the angles between the \emph{z} axis
and vectors ${\bm S}_{1}$ and ${\bm S}_{0}$, respectively, reckoned counterclockwise
from the \emph{z} axis.

\section{Theory}

The above qualitative analysis of orientation of the vectors ${\bm
S}_0, {\bm S}_1$ and the angle between them is based on
Eqs.~(\ref{k0}) valid for the $\Gamma$-point, ${\bm k}=0$. The
values of ${\bm S}_0$ and ${\bm S}_1$ for arbitrary frequencies
$\hbar\omega_{\rm exc}$ and $\hbar\omega_{\rm det}$ can be found
taking into account the interband optical transitions with ${\bm
k} \neq 0$ where the heavy- and light-hole states are mixed. Such
kind of calculations has been performed for unstrained bulk
zinc-blende-lattice semiconductors~\cite{DPoo}, strained bulk
semiconductors with the top of the heavy-hole subband lying above
that of the light-hole subband~\cite{birivch,Subashiev},
quantum-well structures in the approximation of infinitely-high
barriers~\cite{portnoi} and for finite barriers~\cite{marie} (see
also \cite{hermann}), and semiconductor
superlattices~\cite{Subashiev2}.

In this section we will follow the method of Ref.~\cite{birivch}
developed in order to calculate the optical orientation of
electron spins under interband absorption of circularly polarized
light normally incident on a hexagonal crystal, e.g., CdS or CdSe.
In fact, in \cite{birivch} the quasicubic model of the valence
band structure is used: the crystal splitting of the higher two
valence subbands is obtained by applying an effective uniaxial
strain to a crystal of the cubic symmetry with the $\Gamma_8$
valence band. Here we generalize the approach of
Ref.~\cite{birivch} to consider (i) the oblique incidence of the
exciting light on the sample, and (ii) the opposite sign of the
uniaxial strain resulting in the opposite sequence of the split
subbands, now the light-hole subband lies above the heavy-hole
subband.

The spin-polarized electrons are described by the spin-density
matrix $f_{ss'}$, where $s,s' = \pm 1/2$ are the spin indices.
This matrix can be found from the balance matrix equation
\begin{equation}
\left\{ \frac{\partial \hat{f}}{\partial t}\right\}_{\rm
spin.rel.} + \left\{ \frac{\partial \hat{f}}{\partial
t}\right\}_{\rm Larmor} = \dot{\hat{f}}\:,
\end{equation}
where the first and second terms describe the spin relaxation and
Larmor precession of the electron spins, and the right-hand side
term is the generation rate of the electron spin-density matrix.
Four components of the latter matrix can be calculated by using
the general equation
\begin{equation} \label{rhodot}
\dot{f}_{ss'} \propto \sum\limits_{{\bm k} n} \delta[E_c({\bm k})
- E_{v_n}({\bm k}) - \hbar \omega] \sum_j M_{cs, v_n j}({\bm k})
M^*_{cs', v_n j}({\bm k})\:.
\end{equation}
Hereafter we omit common multipliers and use the following
notations: ${\bm k}$ is the electron wave vector, $E_c({\bm k})$
and $E_{v_n}({\bm k})$ are the electron energies in the conduction
band $c$ and the valence subband $v_n (n = \pm)$ given by
\begin{equation} \label{dispersion}
E_c({\bm k}) = E_0 + \frac{\hbar^2 k^2}{2 m_c} \:,\: E_{v_n}({\bm
k}) = A k^2 - \frac{\Delta_c}{2} \pm R \:,
\end{equation}
\[
R = \sqrt{\left( \frac{\Delta_c}{2} \right)^2 + B^2 k^4 + \frac{B
\Delta_c}{2} (3 k_z^2 - k^2) }\:,
\]
$z$ is the uniaxial-strain axis coinciding with the crystal growth
axis, the index $j=1,2$ enumerates degenerate states in the $v_+$
and $v_-$ valence subbands; $m_c$ is the electron effective mass
in the conduction band, $A$ and $B$ are the standard valence-band
parameters entering the Luttinger Hamiltonian taken in the
spherical approximation ($D = \sqrt{3} B$); $\Delta_c$ is the
splitting of the $\Gamma_8$ valence band at ${\bm k} =0$ called
the crystal splitting. In the following we assume a value of
$\Delta_c$ to be positive in which case the indices $v_+$ and
$v_-$ correspond to the light-hole and the heavy-hole subbands,
respectively, and the parameter $E_0$ coincides with the
fundamental band gap $E_g$. The opposite order of the light- and
heavy-hole subbands considered in Ref.~\cite{birivch} is described
by negative values of $\Delta_c$. The matrix elements for
interband electron transitions, $M_{cs, v_n j}({\bm k})$, are
related to the interband matrix elements of the momentum operator
${\hat{\bm p}}$ by
\[
M_{cs, v_n j}({\bm k}) \propto {\bm e} \cdot {\bm p}_{cs, v_n
j}({\bm k})\:,
\]
where ${\bm e}$ is the light polarization unit vector.

Under oblique incidence with the incidence plane containing the axes $x$ and $z$ (and
perpendicular to the $y$ axis) the initial spin ${\bm S}_0$ of the photogenerated
electrons is related to $\dot{\hat{f}}$ by
\begin{equation} \label{spinmatrix}
S_z = \frac12\ \frac{ \dot{f}_{\frac12,\frac12} - \dot{f}_{-\frac12,-\frac12} }{
\dot{f}_{\frac12,\frac12} + \dot{f}_{-\frac12,-\frac12}}\:,\: S_x = \frac{ {\rm Re} \{
\dot{f}_{\frac12,-\frac12} \} }{\dot{f}_{\frac12,\frac12} +
\dot{f}_{-\frac12,-\frac12}}\:.
\end{equation}

One can show that the sum over $j$ in Eq.~(\ref{rhodot}) can be
reduced to
\begin{equation} \label{dud}
\sum_j M_{cs, v_n j}({\bm k}) M^*_{cs', v_n j}({\bm k}) \propto (
\hat{D} \hat{U}_n \hat{D}^{\dag} )_{ss'}\:,
\end{equation}
where $\hat{D}$ is the 2$\times$4 matrix of the interband matrix
elements $\langle \Gamma_6, s |{\bm e} \cdot {\hat{\bm p}}|
\Gamma_8, m \rangle$ $(s = 1/2, - 1/2; m = 3/2, 1/2, - 1/2, -
3/2)$ calculated between the $\Gamma_6$ conduction and $\Gamma_8$
valence Bloch functions:
\[
D_{sm} = \frac{p_{cv}}{\sqrt{6}} \left[ \begin{array}{cccc} - \sqrt{3} e_+ & 2 e_z & e_- & 0 \\
0 & - e_+ & 2 e_z & \sqrt{3} e_-  \end{array} \right]\:,
\]
$e_{\pm} = e_x \pm {\rm i} e_y$, $p_{cv}$ is the interband matrix
element $\langle S| \hat{p}_z | Z \rangle$ with $S$ and $Z$ being
the standard orbital Bloch functions at the $\Gamma$ point,
\[
\hat{U}_n = \frac{{\cal H}({\bm k}) - E_{v_{\bar n}}({\bm k})}{
E_{v_n}({\bm k}) - E_{v_{\bar n}}({\bm k}) } \hspace{5 mm}
(\bar{n} \neq n)\:,
\]
${\cal H}({\bm k})$ is the effective electron Hamiltonian in the
valence band $\Gamma_8$ split into two subbands of the heavy and
light holes.

Since the matrices $\hat{D}, \hat{D}^{\dag}$ and the electron energy spectrum are
independent of the wave-vector azimuth angle $\varphi_{\bm k}$ the Hamiltonian ${\cal
H}({\bm k})$ in $\hat{U}_n$ can be averaged over $\varphi_{\bm k}$ and reduced to a
diagonal matrix. We omit the further details and present the final result for the
generation matrix (\ref{rhodot}) written as a sum of the partial contributions
$\dot{f}_{ss'}^{(n)}$ due to the optical transitions from the subbands $v_n$.

The partial contributions $\dot{f}_{ss'}^{(n)}$ can eventually be presented in the form
\begin{equation} \label{wvxvy}
\dot{f}_{ss'}^{(n)} = W_n \delta_{ss'} - \frac{\rho^0}{2} (V_x^{(n)} \sin{\theta}\
\sigma_x + V_z^{(n)} \cos{\theta}\ \sigma_z)\:,
\end{equation}
where $\rho^0$ is the degree of circular polarization of the light wave in vacuum,
$\theta$ is the refraction angle, $\sigma_x$ and $\sigma_z$ are the Pauli spin
2$\times$2 matrices. The other notations are defined by
\[
W_n = Z_n \frac{t_s^2 + t_p^2}{2} + \frac{n}{4} F(\theta) Z'_n\:,
\]
\[
V_z^{(n)} = t_s t_p(Z_n + n Z'_n)\:,\:V_x^{(n)} = t_st_p ( Z_n -
\frac{n}{2}Z'_n) \:,
\]
\[
F(\theta) = \frac14 (2 t_s^2 - t_p^2 + 3 t_p^2 \cos{2 \theta})\:,
\]
$t_s$ and $t_p$ are the amplitude transmission coefficients of the
$s$- and $p$-polarized light,
\begin{equation} \label{zsubn}
Z_n \propto 4 \pi \int\limits_0^{+\infty} k^2 dk
\int\limits_0^{\pi} \sin{\theta_{\bm k}} d\theta_{\bm k}
\delta[E_c({\bm k}) - E_{v_n}({\bm k}) - \hbar \omega]\:,
\end{equation}
\begin{equation} \label{z'subn}
Z'_n \propto 4 \pi \int\limits_0^{+\infty} k^2 dk
\int\limits_0^{\pi} \sin{\theta_{\bm k}} d\theta_{\bm k}
\frac{Bk^2(1 - 3 \cos^2{\theta_{\bm k}}) - \Delta_c}{R}
\delta[E_c({\bm k}) - E_{v_n}({\bm k}) - \hbar \omega]\:.
\end{equation}
Note that $Z_n$ is proportional to the reduced density of states
``conduction band - valence subband $n$''.

The integrals (\ref{zsubn}) and (\ref{z'subn}) can be readily
calculated assuming \[ \eta = \frac{|B|}{(\hbar^2/2m_c) + |A|} \ll
1 \:.
\]
In GaAs the dimensionless parameter $\eta$ is about 0.23. For the
light propagating in the direction ${\bm n} = (\sin{\theta}, 0,
\cos{\theta})$ one has
\[
W_+ = \sqrt{\chi_+} \left[ \frac{t_s^2 + t_p^2}{2} - \frac12\
\Pi(\chi_+) F(\theta) \right]\:,
\]
\[
W_- = \left\{ \begin{array}{c} \hspace{3 cm} 0 \hspace{2.6 cm}
\mbox{for}
\hspace{0.5 cm} 1 > \varepsilon > 0 \:,\\
\sqrt{\chi_-} \left[ \frac{t_s^2 + t_p^2}{2} + \frac12\
\Pi(\varepsilon_-) F(\theta) \right] \hspace{0.5 cm} \mbox{for}
\hspace{1 cm} \varepsilon > 1\:.
\end{array} \right.
\]
Here
\[
\chi_+ = \eta \varepsilon\:,\: \varepsilon = \frac{\hbar \omega -
E_g}{\Delta_c}\:,
\]
\[
\chi_- = \left\{ \begin{array}{c} \hspace{1 cm} 0 \hspace{1.2 cm}
\mbox{for} \hspace{0.6 cm} 1 > \varepsilon > 0 \:,\\ \eta
(\varepsilon - 1) \hspace{0.5 cm} \mbox{for} \hspace{1 cm}
\varepsilon
> 1\:,
\end{array} \right.
\]
$\omega$ is the light frequency, and the function $\Pi$ of a
variable $X$ is defined by
\begin{equation} \label{Pi}
\Pi(X) = \frac14 \left( \frac{3 + 2 X - 4 X^2}{ \sqrt{6 X}}
\arcsin{\sqrt{\frac{6 X}{1 + 4 X^2 + 2 X}}} + |1 - 2 X |
\right)\:.
\end{equation}
The photoelectron average spin components in the directions $z$
and $x$ are determined by
\begin{eqnarray} \label{szsx}
S_z = \frac12 \rho^0 t_st_p\cos{\theta} \frac{R^+_z + R^-_z}{W_+ +
W_-} \:, \\ S_x = \frac12 \rho^0 t_st_p \sin{\theta} \frac{R^+_x +
R^-_x}{W_+ + W_-}\:, \nonumber
\end{eqnarray}
where
\[
R^+_z = - \frac12 \sqrt{\chi_+}\ [ 1 - 2 \Pi(\chi_+) ]\:,\: R^+_x
= - \frac12 \sqrt{\chi_+}\ [ 1 + \Pi(\chi_+) ]\:,
\]
$R^-_z = R^-_x = 0$ if $1 > \varepsilon > 0$, and
\[
R^-_z =  - \frac12 \sqrt{\chi_-}\ [ 1 + 2 \Pi(\chi_-) ] \:, \:
R^-_x =  - \frac12 \sqrt{\chi_-}\ [ 1 - \Pi(\chi_-) ] \:,
\]
if $\varepsilon > 1$.

The above equations correspond to positive $\Delta_c$. However,
they are also valid for negative $\Delta_c$ if the function
(\ref{Pi}) is replaced by
\[
\Pi(X) = - \frac14 \left( 1 + 2 X + \frac{3 - 2 X - 4 X^2}{\sqrt{6
X}} \ln{\frac{1 + 2 X + \sqrt{6 X}}{\sqrt{1 + 4 X^2 - 2 X}}}
\right)\:.
\]

The angle $\varphi$ between the vector of the average spin and the
axis $z$ equals to
\begin{eqnarray} \label{varphi}
\varphi &&= \arctan{ \left(\frac{R_x}{R_z} \tan{\theta} \right)} +
\left( 1 - {\rm sign}\{ S_z\} \right)\ \frac{\pi}{2} \\
&&\approx  \arctan{ \left( \frac{R_x}{R_z} \theta \right)} +
\left( 1 - {\rm sign}\{ S_z\} \right)\ \frac{\pi}{2}\:, \nonumber
\end{eqnarray}
where $R_z = R_z^+ + R_z^-, R_x = R_x^+ + R_x^-$. Here we take
into account that, in the medium with a big index of refraction,
the refraction angle $\theta$ is small even for remarkable
incidence angles.

Under photoexcitation near the fundamental edge, $\hbar \omega \approx E_g$, one has
$\chi_- = 0$, $W_- = 0$ and the optical transitions are allowed only from the upper
light-hole subband. For $0 < \varepsilon = (\hbar \omega - E_g)/\Delta_c \ll 1$, the
function $\Pi(\chi_+)$ in (\ref{Pi}) tends to 1, and according to Eq.~(\ref{szsx}) the
ratio $S_x/S_z$ is given by $- 2 \tan{ \theta}$ and $\varphi$ by $- \arctan({2
\tan{\theta}})$ in agreement with the first equation~(\ref{shhslh}). At the edge of the
transitions from the heavy-hole subband, $\varepsilon - 1 = (\hbar \omega - E_g -
\Delta_c)/\Delta_c \ll 1$, the function $\Pi(\chi_-) \to 1$ and, therefore, $R^-_x \to
0$ and $R^-_z$ is negative. Therefore, for electrons excited from the top of the $v_-$
subband, $\varphi \to \pi$, in agreement with the second equation (\ref{shhslh}).

\section{Calculation, experimental results and discussion}
\subsection{Specular configuration} It is convenient to use the
reduced excitation and detection energies $\varepsilon_{\rm exc}=(\hbar\omega_{\rm
exc}-E_g)/\Delta_c$ and $\varepsilon_{\rm det}=(\hbar\omega_{\rm det}-E_g)/\Delta_c$.
Our measurements have been made for a large excitation energy $\varepsilon_{\rm
exc}=7.2$ permitting the crystal splitting of the valence subbands to be neglected.
Below we present the results of calculation for that energy. They show that in this case
$\beta\approx180^{\circ}-\theta$, i.e. ${\bm S}_0 \downarrow\uparrow {\bm n}_0$, and
$|\bm S_{0}|=0.25\cos\theta\approx0.25$, which is in full agreement with the above
qualitative analysis based on the expressions (\ref{k0}).

\begin{figure}
\includegraphics[width=0.45\linewidth]{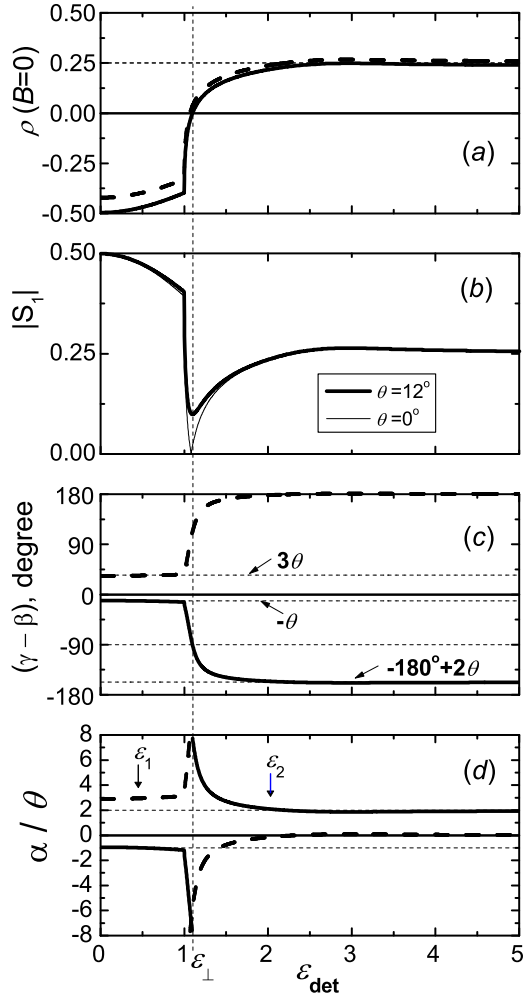}
\caption{\label{alfa&Ro(Edet)} The PL circular polarization
(\emph{a}), the spin $|S_1|$ (\emph{b}), the angles
($\gamma-\beta$) (\emph{c}) and $\alpha$ (\emph{d}) calculated as
a function of the PL detection energy for specular (solid) and
backscattering (dashed) configurations for the refraction angle
$\theta=12^{\circ}$, corresponding to the incidence angle of
48$^{\circ}$, and excitation energy $\varepsilon_{\rm exc}=7.2$ in
the absence of magnetic field. In \emph{b}, the dependence
$|S_1(\varepsilon_{\rm det})|$ is shown also for the normal
incidence, $\theta = 0$. In the case of specular configuration the
normal to the sample surface coincides with the bisectrix between
the exciting beam and the direction of the PL detection. }
\end{figure}

The dependencies of the polarization $\rho(0)=\rho_{0}\cos(\gamma-\beta)$ and of the
angle $(\gamma-\beta)$ on the detection energy $\varepsilon_{\rm det}$ calculated for
zero magnetic field and  specular configuration are shown by solid lines in
Fig.\,\ref{alfa&Ro(Edet)}a and Fig.\,\ref{alfa&Ro(Edet)}c. One can see that for the
detection energy $\varepsilon_{\rm det}=0$, when recombination with the light holes with
$k=0$ is only possible and the vector $\bm{S}_{1}$ is directed at the angle
$\gamma=180^{\circ}-2\theta$ to the $z$ axis, the angle $(\gamma-\beta)=-\theta$ and
polarization $\rho(0)\approx-0.5$. In this case, $|\bm S_{1}|=0.5$. When increasing
$\varepsilon_{\rm det}$, the admixture of the heavy-hole states leads to the increase of
$\rho(0)$ and $(\gamma-\beta)$. These changes of $\rho(0)$ and $(\gamma-\beta)$ are
small when $\varepsilon_{\rm det}$ varies from 0 to 1. However, the $\rho(0)$ and
$(\gamma-\beta)$ dependencies have abrupt humps at $\varepsilon_{\rm det}=1$, when the
transitions from the heavy-hole subband are switched on, involving the creation of
electrons with opposite signs. With further increase of $\varepsilon_{\rm det}$, the
quantity $\rho(0)$ increases rapidly, then reverses its sign from negative to positive
and approaches its maximal value $\rho(0)=0.25$ at $\varepsilon_{\rm det}\approx2$,
while the angle $(\gamma-\beta)$ approaches the maximal value of
$(-180^{\circ}+2\theta)$, corresponding to which are $\gamma=\theta$ and
$|\bm{S}_{1}|=0.25$. Those values of $\rho(0)$ and $(\gamma-\beta)$ remain practically
invariable with $\varepsilon_{\rm det}$ increasing still further.

The angle $(\gamma-\beta)=-90^{\circ}$ ($\bm{S}_{1}\perp\bm{S}_{0}$) when the detection
energy is equal to $\varepsilon_{\perp}$ (shown by a vertical dashed line in
Fig.\,\ref{alfa&Ro(Edet)}). Therefore at such an energy of detection the dependence of
$\rho(0)$ on $\varepsilon_{\rm det}$ passes through zero, although
$|S_{1}(\varepsilon_{\perp})|\neq0$ (see thick solid line in
Fig.\,\ref{alfa&Ro(Edet)}b). Note that $|S_{1}(\varepsilon_{\perp})|=0$ only when the
direction of PL detection coincides with the axis of strain
($\bm{n}_{1}\parallel\bm{\nu}$, $\theta=0$) (thin solid line in
Fig.\,\ref{alfa&Ro(Edet)}b).

Since $\cos(\gamma-\beta)$ and $\sin(\gamma-\beta)$ reverse their signs as the argument
changes by 180$^{\circ}$, the $(\gamma-\beta)$ angle can be substituted by the angle
$\alpha$, which, as shown in Fig.\,\ref{alfa&Ro(Edet)}d, is approximately equal
$-\theta$ at $\varepsilon_{\rm det}<1$, $2\theta$ at $\varepsilon_{\rm det}>2$  and
changes drastically within the range $1<\varepsilon_{\rm det}<2$, making a leap from
-90$^{\circ}$ to +90$^{\circ}$ at $\varepsilon_{\rm det}=\varepsilon_{\perp}$
($\varepsilon_{\perp}\approx1.09$). With such definition of angle $\alpha$,
expression~(\ref{HanleBulk}) for the Hanle effect takes the form:
\begin{equation}
\label{obliqueHanle} \rho(B)=\rho_0\frac{\cos \alpha + \varphi
\sin \alpha}{1+\varphi^2}{\rm sign}(\varepsilon_{\rm
det}-\varepsilon_{\perp})~~,
\end{equation}
where $\rho_0=4|S_0||S_1|T_s/\tau$.

We consider the coupled spin-system of free and bound electrons in the model of
spin-dependent recombination proposed by Weisbuch and Lampel~\cite{WL1974}, applied by
Paget~\cite{Page} and generalized in \cite{JETPLett2005}. In this model the
magnetic-field dependence of the PL polarization can be with high accuracy described by
two Lorentzians:
\begin{equation}
\label{HanleSplitted} \rho(B)=\left[\rho_0\frac{\cos \alpha + \varphi \sin
\alpha}{1+\varphi^2}+\rho_{0c}\frac{\cos \alpha + \varphi_c \sin
\alpha}{1+\varphi_c^2}+\rho_{\rm res}\right] {\rm sign}(\varepsilon_{\rm
det}-\varepsilon_{\perp})~,
\end{equation}
where the first and the second term in the square brackets describe depolarization of
the free and the bound electrons, respectively, $\varphi_{c}=g_{c}\mu_B BT_{sc}/\hbar$
is the angle of rotation of the mean spin of localized electrons $\bm{S_c}$ in a
magnetic field $\bm{B}$ during the lifetime of their spin $T_{sc}$, $\rho_{\rm res}$ is
the residual polarization, observed experimentally in high magnetic field.

It is convenient to measure signs of $g$ and $g_{c}$ at $\varepsilon_{\rm det}\gg 1$ and
$\varepsilon_{\rm det}\ll 1$. As seen from Fig.\,\ref{alfa&Ro(Edet)}d, in the first case
$\alpha=+2\theta$ (as in bulk unstrained semiconductor), and in the second case
$\alpha=-\theta$ .

\begin{figure}
\includegraphics[width=0.4\linewidth]{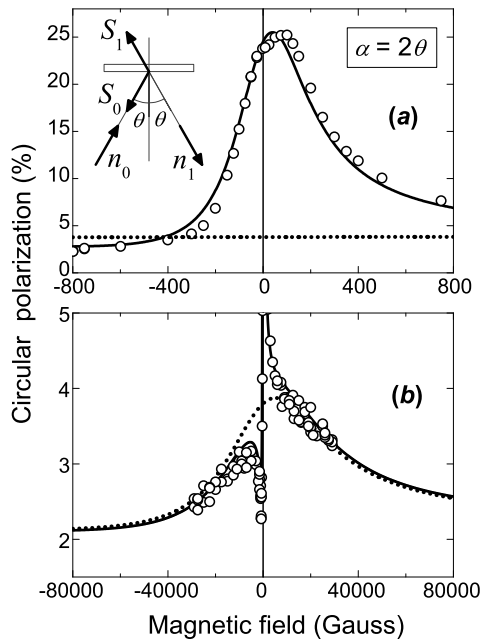}
\caption{\label{2teta} Experimental (circles) and calculated
(solid, dotted) Hanle curves obtained in GaAs$_{0.979}$N$_{0.021}$
in the specular configuration set-up at high-energy excitation and
high-energy recombination when $\alpha\approx2\theta$.
$\hbar\omega_{\rm exc}=1.312$\,eV, $\hbar\omega_{\rm
det}=E_2=1.163$\,eV, $\theta=12^{\circ}$. Inset illustrates
relative orientation of the spins $\bm{S}_0$, $\bm{S}_1$ and the
unit vectors $\bm{n}_0$, $\bm{n}_1$. }
\end{figure}

{\underline {\emph{High energy detection} ($\alpha=+2\theta$)}}

We realized the first case at $\varepsilon_{\rm det}=\varepsilon_{2}=2.03$ (here, just
as in all the following measurements, $\varepsilon_{\rm exc}=7.2$ and the refraction
angle $\theta=12^{\circ}$). The experimental Hanle curve (open circles in
Fig.\,\ref{2teta}) has a manifestly asymmetric shape in this case. The maximum of its
narrow part (Fig.\,\ref{2teta}a), which corresponds to depolarization of bound
electrons, shifts to positive values of the magnetic field. For the employed specular
geometry of experiment, to this corresponds the positive sign of $g_c$
\cite{JETPLett2005} (note that the Hanle curve, measured in the same geometry in bulk
GaAs, in which $g<0$, has a maximum at $B<0$). The maximum of the broad part of the
experimental curve $\rho(B)$ determined by depolarization of free electrons is less
pronounced, since it is superimposed by the depolarization curve of bound electrons,
which has a greater-by-an-order amplitude in zero magnetic field. The polarization of
bound electrons, however, decreases rapidly as the magnetic field increases, and we can
neglect its contribution at $|B|>15$\,kG. The solid curve in Fig.\,\ref{2teta} is
calculated from Eq.\,(\ref{HanleSplitted}) with the same values of depolarization curve
halfwidths for bound and free electrons $B_{1/2}^{c}=185$\,G and $B_{1/2}=25000$\,G that
have been obtained from the fitting of the experimental Hanle curves in
Fig.\,\ref{0teta}, measured under normal excitation and detection (the same values
$B_{1/2}^{c}$ and $B_{1/2}$ will be used below in approximation of experimental Hanle
curves measured under different experimental conditions). One can see that  this curve
approximates the experimental dependence $\rho(B)$ satisfactorily. This makes it
possible to extract the partial Hanle curve for free electrons that is described by the
first term in Eq.\,(\ref{HanleSplitted}) and shown by the dotted line in
Fig.\,\ref{2teta}. As the dotted line has a maximum at $B>0$ (Fig.\,\ref{2teta}b), we
can conclude that $g>0$.

\begin{figure}
\includegraphics[width=0.4\linewidth]{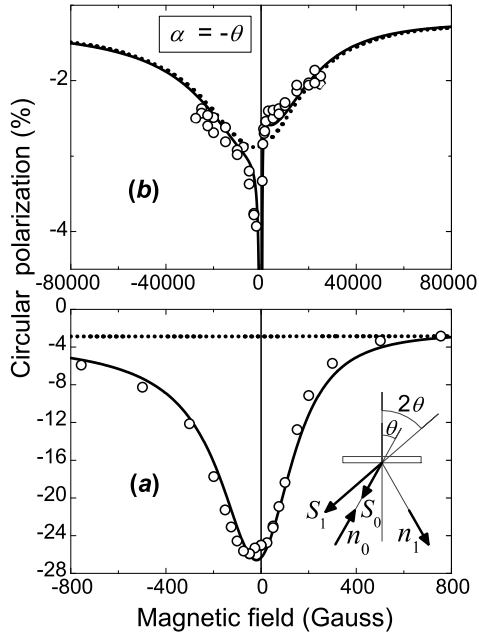}
\caption{\label{(-1)Teta} Experimental (circles) and calculated
(solid, dotted) Hanle curves obtained in GaAs$_{0.979}$N$_{0.021}$
in the specular configuration at high-energy excitation and the
electron recombination with the light holes when $\alpha\approx
-\theta$. $\hbar\omega_{\rm exc}=1.312$\,eV, $\hbar\omega_{\rm
det}=E_1=1.117$\,eV, $\theta=12^{\circ}$. }
\end{figure}

{\underline {\emph{Low energy detection} ($\alpha=-\theta$)}}

The experimental dependence $\rho(B)$ for this case that has been
realized at $\varepsilon_{\rm det}=0.45$, is shown by open circles
in Fig.\,\ref{(-1)Teta}. The negative sign of the measured
luminescence polarization results from the fulfillment of the
condition $\varepsilon_{\rm det}<\varepsilon_{\perp}$. Solid curve
in Fig.\,\ref{(-1)Teta} is calculated using
Eq.\,(\ref{HanleSplitted}) for $\alpha=-\theta$. The maxima (in
the absolute value) on the depolarization curves for the bound
(Fig.\,\ref{(-1)Teta}a) and free (dotted line in
Fig.\,\ref{(-1)Teta}b) electrons are situated at $B<0$, which also
attests to positive signs of $g$ and $g_c$.

\subsection{Backscattering configuration}

\begin{figure}
\includegraphics[width=0.4\linewidth]{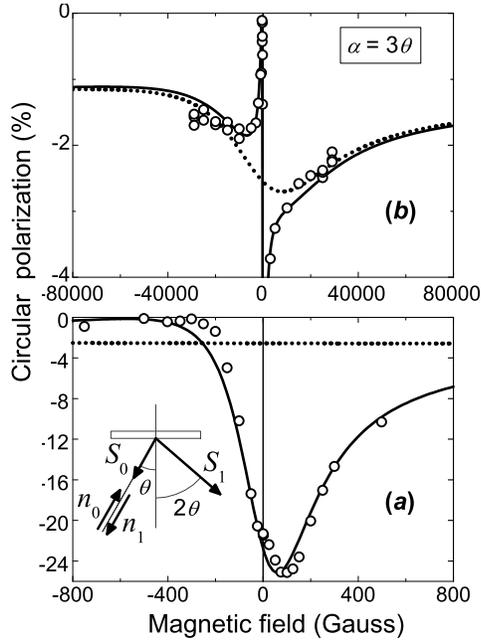}
\caption{\label{3teta} Experimental (circles) and calculated
(solid, dotted) Hanle curves obtained in GaAs$_{0.979}$N$_{0.021}$
in the backscattering configuration at high-energy excitation and
the electron recombination with light holes when $\alpha\approx
3\theta$. $\hbar\omega_{\rm exc}=1.312$\,eV, $\hbar\omega_{\rm
det}=E_1=1.117$\,eV, $\theta=12^{\circ}$. }
\end{figure}

Deformation-induced splitting of the light- and heavy-hole subbands allows the Hanle
curve asymmetry to be observed even in the backscattering configuration where the
directions of detection and excitation are antiparallel,
$\bm{n}_1\downarrow\uparrow\bm{n}_0$. To realize that possibility, the magnetic field
$\bm{B}$, lying in the sample surface plane, is directed perpendicular to the exciting
beam, while the sample is turned around vector $\bm{B}$ in such a way that there is
nonzero angle $\theta$ between the normal to the sample (axis $\bm{\nu}$) and the
exciting beam (vector $\bm{n}_0$), as shown in the insert in Fig.\,\ref{3teta}.

The dependence of angle $\alpha$ between vectors $\bm{S}_0$ and
$\bm{S}_1$ on detection energy $\varepsilon_{\rm det}$ calculated
for backscattering configuration and $\varepsilon_{\rm exc}=7.2$
is shown by dashed line in Fig.\,\ref{alfa&Ro(Edet)}d. One can see
that $\alpha=+3\theta$ if $\varepsilon_{\rm det}<1$ and $\alpha
\approx 0$ if $\varepsilon_{\rm det}>2$.

{\underline {\emph{Low energy detection} ($\alpha=+3\theta$)}}.

At $\varepsilon_{\rm det}<1$ angle $\alpha\approx3\theta$, which makes the asymmetry of
the experimental Hanle curve (open circles in Fig.\,\ref{3teta}) more pronounced than in
the case of the specular configuration when $\alpha\approx-\theta$. The negative sign of
polarization is due to the fact that $\varepsilon_{\rm
det}<\varepsilon_{\perp}\approx1.07$. Solid and dotted curves in Fig.\,\ref{3teta} are
calculated from Eq.\,(\ref{HanleSplitted}) and the second term of
Eq.\,(\ref{HanleSplitted}), respectively, for $\alpha=3\theta$. It is seen that the
maxima of depolarization curves both for the bound (Fig.\,\ref{3teta}a) and the free
electrons (dotted line in Fig.\,\ref{3teta}a) are found in the region of positive values
of the magnetic field. For a backscattering configuration this indicates that the signs
of $g$ and $g_c$, which is in full agreement with the result obtained with the use of
specular configuration.

{\underline {\emph{High energy detection} ($\alpha=0$)}}.

\begin{figure}
\includegraphics[width=0.4\linewidth]{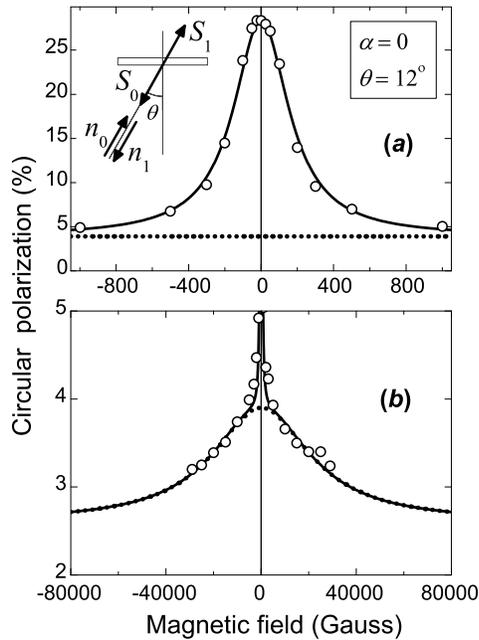}
\caption{\label{0alfa-12teta} Experimental (circles) and
calculated (solid, dotted) Hanle curves for
GaAs$_{0.979}$N$_{0.021}$ in the case of backscattering
configuration at high-energy excitation and high-energy
recombination when $\alpha=0$. $\hbar\omega_{\rm exc}=1.312$\,eV,
$\hbar\omega_{\rm det}=E_{2}=1.163$\,eV, $\theta=12^{\circ}$. }
\end{figure}

At $\varepsilon_{\rm det}>2$ (when deformation splitting can be
neglected) the angle $\alpha \approx 0$ and, according to
Eq.\,(\ref{HanleSplitted}), the Hanle curve must be symmetrical
and insensitive to the sign of \emph{g}-factor. Indeed, the
experimental Hanle (open circles in Fig.\,\ref{0alfa-12teta}),
measured at $\varepsilon_{\rm det}=\varepsilon_{2}=2.03$, is
symmetric.

The experimental dependencies $\rho(B)$ presented above have been measured under
high-energy excitation ($\varepsilon_{\rm exc}\gg 1$) and detection energy
$\varepsilon_{\rm det}\gg 1$ or $\varepsilon_{\rm det}<1$, when angle $\alpha$ remains
practically invariable with varying $\varepsilon_{\rm det}$. Under these conditions the
greatest absolute value of the angle $\alpha$ and, respectively, the greatest asymmetry
of Hanle curve are observed in specular configuration at $\varepsilon_{\rm det}\gg 1$
($\alpha=2\theta$) and in backscattering configuration at $\varepsilon_{\rm det}<1$
($\alpha=3\theta$). A greater asymmetry of Hanle curve can be obtained in specular
configuration under excitation and detection with only a light hole participating
($\varepsilon_{\rm det}<\varepsilon_{\rm exc}<1$), when $\alpha=-4\theta$
\cite{FTT1997}. Note that the maximum asymmetry of Hanle curve can be realized both in
specular and backscattering configurations under high-energy excitation and detection
energy $\varepsilon_{\rm det}=\varepsilon_{\perp}$, when in the absence of magnetic
field the vectors $\bm{S}_1$ and $\bm{S}_0$ are perpendicular to each other (angle
$|\gamma-\beta|=90^{\circ}$) and $\rho\propto\bm{S_0}\bm{S_1}=0$, while in a magnetic
field $\rho\propto\pm|S_0| |S_1| \varphi/(1+\varphi^2)$, where the sign +($-$) before
the right-hand side corresponds to the specular (backscattering) configuration. However,
at $\varepsilon_{\rm det}=\varepsilon_{\perp}$ the modulus of $\bm{S}_1$ vector
diminishes sharply and is near zero if the refraction angle does not exceed a few
degrees ($|S_1(\varepsilon_{\perp})|=0$ for $\theta=0$), amounting to $\approx0.1$ only
at angles $\theta$ close to the maximum value ($\theta_{\rm max}=16^{\circ}$ in GaAs
\cite{note1}) (see Fig.\,\ref{alfa&Ro(Edet)}b).

Experimental dependencies $\rho(B)$ measured in GaAsN consist of narrow and broad parts.
The large amplitude of the narrow part bears witness to strong polarization of bound
electrons, arising due to spin-dependent recombination. We have approximated the
experimental curves $\rho(B)$ by sum of two Lorentzians, one of which describes the
depolarization of bound electrons, and the other, that of free electrons. At the same
time, we have shown in Ref.\,\cite{JETPLett2005} that the polarizations of free and
bound electrons are summed up additively ($\rho\propto S+S_c$) only in the limit of low
polarization of bound electrons. If polarization of bound electrons is strong, it
becomes necessary to solve a system of nonlinear equations, coupling polarizations of
free and bound electrons \cite{JETPLett2005}. As the preliminary analysis shows, in that
case a substantially better fit of the experimental Hanle curve is obtained for the
region of low magnetic fields, where the polarization of bound electrons dominates. The
detailed theoretical description of the compound Hanle effect under an arbitrary
polarization (100\% including) of bound electron will be presented in a separate
publication. Here we note that the shape of Hanle curves (in particular their
asymmetry), obtained from the solution of a system of nonlinear equations does not
differ qualitatively from the results of approximation by two Lorentzians, which makes
it possible to employ the latter for determination of the signs of $g$ and $g_c$.

\section{Summary}
In order to determine the sign of the electron $g$ factor in a
semiconductor film, we have applied the method based on analysis
of the asymmetrical Hanle effect under oblique incidence of the
exciting light onto the sample and detection of the
photoluminescence at oblique emission. The relationship between
the $g$ factor sign and direction of the asymmetrical shift of the
Hanle curve is governed by the valence band structure,
particularly, by its splitting due to an internal uniaxial strain
in the film, and depends on the photoexcitation and photodetection
energies. In the present experimental and theoretical work we have
extended the method to the case where the internal strain shifts
the light-hole subband upwards relative to the heavy-hole subband.
The interband optical orientation of electron spins, circular
polarization of photoluminescence, and shape of the Hanle curve
have been calculated in dependence on the energies of excitation
and detection as well as on the angles of incidence and secondary
emission.

The experimental study has been performed on uniaxially compressed GaAs$_{1-x}$N$_x$
epitaxial films. The observed Hanle effect is a result of the magnetic-field induced
depolarization of the coupled system of spin-polarized free electrons and electrons
bound on deep paramagnetic centers. The asymmetry of the resultant Hanle curve has been
measured at room temperature under oblique incidence of the pump radiation and detection
in the specular or backscattering configurations. Positive signs of the gyromagnetic
$g$-factors of the free and bound electrons in the GaAs$_{0.979}$N$_{0.021}$ alloy have
been deduced from comparison between experiment and theory.

\ack We are grateful to late B.P. Zakharchenya for stimulating interest in this work and
to M.M.~Glazov, K.V.~Kavokin and V.M.~Ustinov for fruitful discussions. Partial support
by the Russian Foundation for Basic Research, grants of the Russian Academy of Sciences,
and TARA project of the University of Tsukuba is acknowledged.


\begin{thebibliography}{10}
\bibitem{note1} To be precise, $\tan\zeta=2\tan\theta$, where $\zeta$ is the angle
between $\bm{S}_{0lh}$ and $\bm\nu$. Since the angle $\theta$ between $\bm{n}_0$ and
$\bm\nu$ cannot exceed 16$^{\circ}$ because of the large refractive index, approximately
3.6, of GaAs and GaAsN, we can accept $\zeta\approx2\theta$.
\end{thebibliography}
\end{document}